\documentclass[twocolumn,showkeys,showpacs,pre]{revtex4}
\usepackage{amsfonts, amsmath, latexsym}
\usepackage[dvips]{graphicx}

\newtheorem{1}{Theorem}

\begin{document}

\title{Entropic effects in large-scale Monte Carlo simulations}

\author{Cristian Predescu} 
\email{cpredescu@comcast.net}

\affiliation{
Department of Chemistry and Kenneth S. Pitzer Center for Theoretical Chemistry, University of California, Berkeley, California 94720
}
\date{\today}
\begin{abstract}
The efficiency of Monte Carlo samplers is dictated not only by energetic effects, such as large barriers, but also by entropic effects that are due to the sheer volume that is sampled. The latter effects appear in the form of an entropic mismatch or divergence between the direct and reverse trial moves. We provide lower and upper bounds for the average acceptance probability in terms of the R\'enyi divergence of order $1/2$. We show that  the asymptotic finitude of the entropic divergence is the necessary and sufficient condition for non-vanishing acceptance probabilities in the limit of large dimension.  Furthermore, we demonstrate that the upper bound is reasonably tight by showing that the exponent is asymptotically exact for systems made up of a large number of independent and identically distributed subsystems. For the last statement, we provide an alternative proof that relies on the reformulation of the acceptance probability as a large deviation problem. The reformulation  also leads to a class of low-variance estimators for strongly asymmetric distributions. We show that the entropy divergence causes a decay in the average displacements with the number of dimensions $n$ that are simultaneously updated. For systems that have a well-defined thermodynamic limit, the decay is demonstrated to be $n^{-1/2}$ for random-walk Monte Carlo and $n^{-1/6}$ for Smart Monte Carlo (SMC). Numerical simulations of the $\mathrm{LJ}_{38}$ cluster show that SMC is virtually as efficient as the Markov chain implementation of the Gibbs sampler, which is normally utilized for Lennard-Jones clusters. An application of the entropic inequalities to the parallel tempering method demonstrates that the number of replicas increases as the square root of the heat capacity of the system. 
\end{abstract}
\maketitle

\section{Introduction}

The Markov Chain Monte Carlo algorithms \cite{Kal86, Liu01}  sample a configuration or phase space by means of an ergodic Markov chain that leaves the equilibrium distribution invariant. Except for a few remarkable exceptions, such as Swendsen and Wang's cluster algorithm and follow-ups \cite{Swe87, Kan88, Mak07}, the chain's transition matrix is constructed by means of the rejection technique, according to the Metropolis-Hastings prescription.  Typically, the ensuing Markov chains are local in nature. In the limit of small displacements (nearest-neighbor moves on a lattice), the generated stochastic dynamics becomes an activated diffusive process, akin to the Smoluchowski dynamics. For such dynamics,  the rate of equilibration is controlled by the mean escape times from the dominant basins, which times are proportional to $\exp(+\beta \Delta E)$, with $\beta$ being the inverse temperature and $\Delta E$ being the energy barrier. The transition state theory can provide more accurate estimates, which involve free energy barriers. 

To attenuate the negative influence of the energetic barriers, one may attempt to alter the sampled distribution by means of carefully-chosen weighting factors, as is the purpose of umbrella sampling \cite{Tor74} and multicanonical ensembles \cite{Ber92}. If such modifications are not what is desired, the mixing times can be improved by attempting larger jumps, ideally, directly through a barrier. However, examples of truly global algorithms are few and their existence depends on specific properties of the systems that are investigated. General-purpose algorithms that achieve non-locality, such as, J-walking \cite{Fra90}, simulated tempering \cite{Mar92},  or replica exchange \cite{Swe86, Gey91, Liu01}, do so by means of extending the dimensionality of the system, therefore, introducing new parameters that provide alternative communication routes. In the enlarged space, the algorithms remain local, making the following question pertinent to the design of algorithms: What determines the size of the jumps? In this paper, we  provide the answer in the form of an entropic mismatch between the distribution that is sampled and the trial distribution utilized in the construction of the Monte Carlo transition matrix. 

For standard random walk Metropolis samplers, numerical observations evidencing the entropic nature of the displacement-limiting factors exist in the form of experimental knowledge that updating many particles simultaneously results in a decrease of the average displacements.  Nevertheless, somewhat at odds with the observed dimensional sensitivity, the standard justification for the decrease has been that the likelihood that the particles collide is increased. Owing to the ensuing increase in the energy, high-dimensional moves are more likely to be rejected.  It was recently realized that this explanation is erroneous and that the influence of the dimensionality on the efficiency of samplers is an entropic effect \cite{Pre05}. Numerical experiments have shown that the effect features a behavior similar to that predicted for systems made up of independent particles, which particles cannot collide. In the original work, the effect has not been properly quantified. Given the importance of the rejection-based Monte Carlo algorithms in the physical sciences, the author feels that a proper analysis is desirable. More generally, acquiring tools necessary to quantify the effect of dimensionality is of definite importance, as Monte Carlo methods are techniques intended to cope specifically with large dimensional systems.

As mathematical technique, the point of start are some simple yet strong inequalities. Consider a $d$-dimensional system described by a probability density $\rho(\mathbf{x})$, which we shall cast in the normalized Boltzmann form $e^{-\beta V(\mathbf{x})}/Q(\beta)$. The Metropolis-Hastings sampler utilizes the trial distribution $T(\mathbf{y}|\mathbf{x})$. The average acceptance probability is given by the formula
\begin{equation}
\label{eq:I1}
\mathcal{A} = \int \! d\mathbf{x} \! \int \! d\mathbf{y}  \rho(\mathbf{x})T(\mathbf{y}|\mathbf{x})\min\left\{1, \frac{\rho(\mathbf{y})T(\mathbf{x}| \mathbf{y})}{\rho(\mathbf{x})T(\mathbf{y} | \mathbf{x})}\right\}.
\end{equation}
In Section~II, we show that
 \begin{equation}
\label{eq:I3}
\frac{1}{2}\exp\left(-D_{1/2}\right) \leq \mathcal{A} \leq \exp\left(-\frac{1}{2}D_{1/2}\right).
\end{equation}
Here, $D_{1/2} \geq 0$ is the R\'enyi entropic divergence of order $1/2$ defined by the expression \cite{Ren61}
\begin{equation}
\label{eq:I4}
D_{1/2} = - 2\log\left(\int \! d\mathbf{x} \! \int \! d\mathbf{y} \left[\rho(\mathbf{x})T(\mathbf{y}|\mathbf{x})\rho(\mathbf{y})T(\mathbf{x} | \mathbf{y})\right]^{1/2}  \right).
\end{equation}

 The divergence $D_{1/2}$ measures the mismatch between the direct sampling distribution $\rho(\mathbf{x})T(\mathbf{y}|\mathbf{x})$ and the reverse distribution $\rho(\mathbf{y})T(\mathbf{x}|\mathbf{y})$. It is zero if and only if detailed balance is already achieved, an unlikely scenario in practical applications. As an extensive entropy, $D_{1/2}$ is additive for independent probability distributions. As such, if we sample together a system made up of $n$ independent replicas from the product trial distribution $T(\mathbf{y}_1|\mathbf{x}_1)\cdots T(\mathbf{y}_n|\mathbf{x}_n)$, then the acceptance probability $\mathcal{A}_n$ satisfies the inequality
\begin{equation}
\label{eq:I5}
\mathcal{A}_n \leq \exp\left(-\frac{n}{2}D_{1/2}\right).
\end{equation}
Such an exponential decrease has been first observed by Kennedy and Pendleton \cite{Ken91} following an analysis of the performance of Hybrid Monte Carlo simulations for harmonic oscillators. These authors demonstrated that the acceptance probability decreases exponentially unless the time step for the molecular dynamics integrator is suitably decreased as a fractional power of the number of degrees of freedom. To wit in view of Eq.~(\ref{eq:I5}), the choice of harmonic oscillators was fortunate, as their Boltzmann distribution is factorizable in the normal mode frame. We shall rederive Kennedy and Pendleton's results shortly, in other contexts. Nevertheless, there is nothing special about harmonic oscillators except for the guaranteed extensivity of thermodynamic properties, which is, as advocated here, the real penalty source.  

We have already presented some of the mathematical results so that to entice the reader's attention. The exponential decay predicted by Eq.~(\ref{eq:I5}) automatically implies that the entropic divergence $D_{1/2}$ cannot be left unchanged. It must be decreased at least as fast as $1/n$. This can be achieved by tuning some parameters. For example, we may  decrease the average displacements for the random-walk Metropolis algorithm or the ratio between consecutive temperatures for parallel tempering. Details are left for Section~IV. They prove to be consistent with prior analysis, wherever such analysis is available. In particular, the results for the random walk Metropolis algorithm and the Smart Monte Carlo method agree with those obtained by Roberts and coworkers \cite{Rob97, Rob98, Chr05}, whereas the results for parallel tempering simulations agree with those of Kofke \cite{Kof02} and Predescu \emph{et al} \cite{Pre04}.  

A great deal of care is dedicated to the systems made up of a large number $n$ of identical and independent subsystems. Indeed, an analysis of the dependence with the dimensionality requires some form of homogeneity of the system in the thermodynamic limit. In addition, the assumption of independence is a model for the physical reality that sufficiently large subsystems interact weakly. More precisely,  various properties  become extensive in the thermodynamic limit because the contribution of the frontier interactions to these properties becomes comparatively small. In Section~II, we demonstrate that the inequality given by Eq.~(\ref{eq:I5}) is reasonably tight. More precisely, if the trial distribution $T(\mathbf{y}| \mathbf{x})$ utilized for sampling a subsystem is left unchanged as the dimensionality of the overall system is increased, then
\begin{equation}
\label{eq:I6}
-\lim_{n \to \infty} n^{-1}\log\left(\mathcal{A}_n \right) = \frac{1}{2}D_{1/2}.
\end{equation}
This last identity is the correct statement for the bad sampling theorem of Ref.~\onlinecite{Pre05}, which incorrectly asserted the law
\begin{equation}
\label{eq:I7}
-\lim_{n \to \infty} n^{-1}\log\left(\mathcal{A}_n \right) = D_1,
\end{equation}
with $D_1$ being the relative Shannon entropy
\begin{equation}
\label{eq:I8}
D_1 = - \int \! d\mathbf{x} \! \int \! d\mathbf{y}\rho(\mathbf{x})T(\mathbf{y}|\mathbf{x}) \log\left[\frac{\rho(\mathbf{y})T(\mathbf{x}| \mathbf{y})}{\rho(\mathbf{x})T(\mathbf{y} | \mathbf{x})}\right].
\end{equation}
Save for this correction, all predictions made in Ref.~\onlinecite{Pre05} with respect to the behavior of the average displacements as a function of dimensionality remain the same. The reason is that the asymptotic finitude of the relative Shannon entropy $D_1$ (also called the Kullback--Leibler divergence or the R\'enyi divergence of order $1$) is a sufficient condition for non-vanishing acceptance  probabilities. 

In Section~III, we recast the evaluation of the average acceptance probability as a large deviation problem. We then utilize Cram\'er's large deviation theorem to provide a second demonstration of Eq.~(\ref{eq:I6}). The rare events associated with large deviation problems are a source of exponentially increasing variances for quantities such as acceptance probabilities or ratios of partition functions. To alleviate the overlap problem, a class of low-variance estimators for strongly asymmetric distributions is introduced, along with a demonstration of their statistical efficiency.

In Section~V, we find that the Smart Monte Carlo method of Rossky, Doll, and Friedman \cite{Ros78} features an improved theoretical scaling of $n^{-1/6}$. Numerical verification recovers the predicted scaling for correlated systems, namely embedded Lennard-Jones clusters. The findings are in agreement with the results of Roberts and Rosenthal \cite{Rob98}, who have obtained the same $n^{-1/6}$ scaling for systems made up of statistically independent subsystems.  A simulation of the LJ$_{38}$ cluster shows that Smart Monte Carlo is roughly as efficient as the Metropolis algorithm with single-particle updates, which is the technique normally utilized for clusters. For more expensive potentials, it is argued that Smart Monte Carlo will likely be more efficient provided that the number of atoms updated simultaneously is in the range of tens to  hundreds.

Section~VI contains the conclusions of the paper, which are mostly recommendations on the design of Monte Carlo algorithms so that to minimize the negative impact of the entropic effects. 

\section{Proof of the mathematical claims}
For the reminder of this paper, the notation $\pi(\mathbf{x}, \mathbf{y}) = \rho(\mathbf{x})T(\mathbf{y}|\mathbf{x})$  turns out to be convenient. Notice that the trial distribution $\pi(\mathbf{x}, \mathbf{y})$ is normally strongly asymmetric and that its symmetry is equivalent to the detailed balance condition. In addition, we shall also need the antisymmetric function 
\begin{equation}
\label{eq:II1}
X(\mathbf{x}, \mathbf{y}) = \log\left[\pi(\mathbf{y}, \mathbf{x})/\pi(\mathbf{x}, \mathbf{y})\right],
\end{equation}
which we regard as a random variable with respect to the probability distribution $\pi(\mathbf{x}, \mathbf{y})$. The expected value of some random variable $Y$, that is, the quantity
\begin{equation}
\label{eq:IIb1}
\int \int d\mathbf{x}d\mathbf{y} \pi(\mathbf{x},\mathbf{y}) Y(\mathbf{x}, \mathbf{y}),
\end{equation}
 will be denoted by $\mathbb{E}(Y)$ or simply $\mathbb{E}Y$ in order to avoid the more cumbersome integral notation. 

 The average acceptance probability reads
\begin{equation}
\label{eq:IIa2}
\mathcal{A} = \mathbb{E} \min\left\{1, e^X\right\} = \mathbb{E} e^{X/2}\min\left\{e^{-X/2}, e^{X/2}\right\}. 
\end{equation}
Observe the equality
\begin{equation}
\label{eq:IIa3}
\min\left\{e^{-X/2}, e^{X/2}\right\} = e^{-|X|/2} \leq 1.
\end{equation} 
It follows that
\begin{equation}
\label{eq:IIa4}
\mathcal{A} = \mathbb{E} e^{X/2} e^{-|X|/2} \leq \mathbb{E} e^{X/2} = e^{-\frac{1}{2}D_{1/2}},
\end{equation}
which is the  upper-bound inequality in Eq.~(\ref{eq:I3}). 

To establish the other inequality, recall that $1/x$ is a convex function on $x \in [0, \infty)$. By virtue of Jensen's inequality, we have
\begin{equation}
\label{eq:IIa5}
\mathcal{A} = \mathbb{E} e^{X/2} \frac{\mathbb{E}e^{X/2} e^{-|X|/2}}{\mathbb{E} e^{X/2}}  \geq \mathbb{E} e^{X/2}  \left(\frac{\mathbb{E} e^{X/2} e^{|X|/2}}{\mathbb{E} e^{X/2}}\right)^{-1}. 
\end{equation}
A helpful identity is 
\begin{equation}
\label{eq:IIb5}
\mathbb{E} e^{(X + |X|)/2} = \mathbb{E} \max\left\{1, e^{X}\right\} \leq \mathbb{E} \left(1+e^X\right) = 2.
\end{equation}
Combining with Eq.~(\ref{eq:IIa5}), we obtain
\begin{equation}
\label{eq:IIa6} 
\mathcal{A}  \geq  \left(\mathbb{E} e^{X/2}\right)^2 \left/\mathbb{E} e^{(X+|X|)/2}\right. \geq \frac{1}{2} \left(\mathbb{E} e^{X/2}\right)^2 = \frac{1}{2}e^{-D_{1/2}}, 
\end{equation}
which is the lower-bound inequality in Eq.~(\ref{eq:I3}). 

Given their relevance for Monte Carlo algorithms, it is worthwhile to recall the definitional features of the R\'enyi divergences. For $\alpha \geq 0$, the R\'enyi divergence of order $\alpha$ for continuous distributions $p(\mathbf{x})$ and $q(\mathbf{x})$ is given by the expression \cite{Ren61}
\begin{equation}
\label{eq:II11}
D_\alpha(p \| q) = \frac{1}{\alpha -1} \log\left[\int  p(\mathbf{x})^\alpha q(\mathbf{x})^{1-\alpha}d\mathbf{x}\right].
\end{equation}
For $\alpha = 1$, the R\'enyi divergence is defined by continuity, as the limit $\alpha \to 1$. The result, which can be obtained by means of l'H\^opital's rule, is the Kullback--Leibler divergence (or the relative Shannon entropy)
\begin{equation}
\label{eq:II12}
D_1(p \| q) = -\int p(\mathbf{x}) \log\left[q(\mathbf{x})/p(\mathbf{x})\right]d\mathbf{x}.
\end{equation}
The R\'enyi divergences are non-negative quantities. They are zero if and only if $p(\mathbf{x})$ and $q(\mathbf{x})$ are identical except for a set of measure zero. As expected of entropies, the R\'enyi divergences are additive for independent distributions. More precisely, if $p(\mathbf{x}, \mathbf{x}') = p_1(\mathbf{x})p_2(\mathbf{x}')$ and $q(\mathbf{x},\mathbf{x}') = q_1(\mathbf{x})q_2(\mathbf{x}')$ then
\begin{equation}
\label{eq:II13}
D_\alpha(p \| q) = D_\alpha(p_1 \| q_1) + D_\alpha(p_2 \| q_2), 
\end{equation}
a relation we shall utilize in the following section. 

As far as the theory developed in the present paper is concerned, of importance is the R\'enyi divergence of order $1/2$, which provides a measure of the mismatch between the direct $\rho(\mathbf{x})T(\mathbf{y}|\mathbf{x})$ and the reverse $\rho(\mathbf{y})T(\mathbf{x}|\mathbf{y})$ sampling probabilities. In words, the divergence quantifies the lack of detailed balance in a way that relates directly to the values of the acceptance probabilities, by virtue of the inequalities we have established. Notice that the divergence of order $1/2$ is symmetric under the permutation of $p$ and $q$. The common value is bounded from above by either of the Kullback--Leibler divergences 
\begin{equation}
\label{eq:II13a}
D_{1/2}(p \| q) = D_{1/2}(q \| p)\leq \min\left\{D_1(p \| q), D_1(q \| p)\right\}. 
\end{equation}
This follows from Jensen's inequality and the convexity of the function $-\log(x)$. 

A different lower bound for the acceptance probability helps demonstrate the limiting result given by Eq.~(\ref{eq:I6}), for independent systems. Let us denote by $\mathbb{E}_s(Y)$ the expected value of some random variable against the symmetric probability distribution $\pi_s(\mathbf{x}, \mathbf{y})$ defined by the expression
\begin{equation}
\label{eq:IIa14}
\frac{\pi(\mathbf{x},\mathbf{y})\exp[X(\mathbf{x}, \mathbf{y})/2]}{\mathbb{E}\exp(X/2)} = \left[e^{D_{1/2}}\pi(\mathbf{x},\mathbf{y})\pi(\mathbf{y},\mathbf{x})\right]^{1/2}. 
\end{equation}
With this notation, the expression for the acceptance probability [see the left-hand side of Eq.~(\ref{eq:IIa5})] becomes 
\begin{equation}
\label{eq:IIa15}
\mathcal{A} = \left(\mathbb{E} e^{X/2}\right) \left(\mathbb{E}_s e^{-|X|/2}\right) = e^{-\frac{1}{2}D_{1/2}} \mathbb{E}_s e^{-|X|/2}.
\end{equation}
Since $e^{-x}$ is a convex function, Jensen's inequality produces
\begin{equation}
\label{eq:IIa16}
\mathcal{A} \geq  e^{-\frac{1}{2}D_{1/2}}\exp\left(-\mathbb{E}_s|X| /2\right).
\end{equation} 
Cauchy's inequality  implies $\mathbb{E}_s|X| \leq (\mathbb{E}_sX^2)^{1/2}$ and so,
\begin{equation}
\label{eq:IIa17}
\mathcal{A} \geq \exp\left[-\frac{1}{2}D_{1/2} - \frac{1}{2}\left(\mathbb{E}_sX^2\right)^{1/2}\right].
\end{equation}

It is worth noting that $(\mathbb{E}_sX^2)^{1/2}$ is in fact a standard deviation. The function $X(\mathbf{x},\mathbf{y})$ is antisymmetric and, consequently, its expectation against the symmetric measure $\pi_s(\mathbf{x}, \mathbf{y})$ is zero. To see the relevance of this observation, assume that we sample together a large number $n$ of identical and statistically independent systems described by the same probability distribution $\rho(\mathbf{x})$. Assume also that we attempt to utilize a same trial distribution $T(\mathbf{y}|\mathbf{x})$, independent of $n$. The formula for the average acceptance probability reads 
\begin{eqnarray}
\label{eq:II5} \nonumber
\mathcal{A}_n = \int \cdots \int d\mathbf{x}_1d\mathbf{y}_1 \cdots d\mathbf{x}_n d\mathbf{y}_n \pi(\mathbf{x}_1, \mathbf{y}_1) \cdots\\ \times \pi(\mathbf{x}_n, \mathbf{y}_n)
\min\left\{1, \prod_{i = 1}^n \frac{\pi(\mathbf{y}_i, \mathbf{x}_i)}{\pi(\mathbf{x}_i, \mathbf{y}_i)}\right\},  
\end{eqnarray}
or
\begin{equation}
\label{eq:IIb6}
\mathcal{A}_n = \mathbb{E}e^{S_n/2} \cdot \mathbb{E}_s e^{ - |S_n|/2}.  
\end{equation}
For the last identity, we have employed Eq.~(\ref{eq:IIa15}), with  $S_n$ defined accordingly by
\begin{equation}
\label{eq:II7}
S_n(\mathbf{x}_1, \mathbf{y}_1, \ldots, \mathbf{x}_n, \mathbf{y}_n) = \log\left[\prod_{i=1}^n\frac{\pi(\mathbf{y}_i, \mathbf{x}_i)}{\pi(\mathbf{x}_i, \mathbf{y}_i)}\right].
\end{equation}
The identity
\begin{equation}
\label{eq:II8}
S_n =\sum_{i=1}^n\log\left[\frac{\pi(\mathbf{y}_i, \mathbf{x}_i)}{\pi(\mathbf{x}_i, \mathbf{y}_i)}\right] = \sum_{i=1}^n X(\mathbf{x}_i, \mathbf{y}_i)
\end{equation}
shows that $S_n$ is a sum of independent and identically distributed random variables $X_i$.
By independence, identical distribution, and the vanishing expectation of each $X_i$, 
\begin{equation}
\label{eq:IIb9}
\mathbb{E}_s S_n^2 =  \sum_{i,j=1}^n \mathbb{E}_s X_iX_j = n\mathbb{E}_s X^2. 
\end{equation}

On the other hand, the additivity of the R\'enyi divergence of order $1/2$ for independent distributions implies
\begin{equation}
\label{eq:IIb7}
\mathbb{E}e^{-S_n/2} = e^{-(n/2)D_{1/2}}.
\end{equation}
We obtain
\begin{equation}
\label{eq:IIb10}
e^{-(n/2)D_{1/2}} \geq \mathcal{A}_n \geq e^{-(n/2)D_{1/2} - (n\mathbb{E}_sX^2)^{1/2}/2}
\end{equation}
and conclude
\begin{equation}
\label{eq:IIb11}
\frac{1}{2}D_{1/2} \leq -\frac{1}{n}\log(\mathcal{A}_n) \leq \frac{1}{2}D_{1/2} + \frac{1}{2n^{1/2}}(\mathbb{E}_sX^2)^{1/2}.  
\end{equation}
This sequence of inequalities produces  Eq.~(\ref{eq:I6}) upon letting $n \to \infty$. We shall construct another proof for Eq.~(\ref{eq:I6}) in the next section, by means of Cram\'er's large deviation theorem. 

\section{The acceptance probability as a large deviation problem}

The basic task of this section is to reformulate the evaluation of the acceptance probability as a large deviation problem. By doing so, we obtain a better understanding of the source of the exponential decrease in the acceptance probability for independent systems. We will find that the source is the rare-event sampling normally associated with large deviation problems. In fact, we shall construct a different proof for Eq.~(\ref{eq:I6}) by means of Cram\'er's large deviation theorem \cite{Dur96}, which quantifies the frequency of such rare events. In addition, we are led to the consideration of special low-variance estimators for the acceptance probability, which can be generalized to other scenarios, as done in Appendix~I.

We start with the following formula for the acceptance probability
\begin{equation}
\label{eq:II3}
\mathcal{A} = \int \int d\mathbf{x} d\mathbf{y} \min \left\{\pi(\mathbf{x}, \mathbf{y}),  \pi(\mathbf{y}, \mathbf{x})\right\}.
\end{equation}
Let $I_A(\mathbf{x}, \mathbf{y})$ be the indicator function of the set
\begin{equation}
\label{eq:II4}
A = \left\{(\mathbf{x}, \mathbf{y}): \pi(\mathbf{x}, \mathbf{y}) <  \pi(\mathbf{y}, \mathbf{x})\right\},
\end{equation}
and let $I_{A^c}(\mathbf{x}, \mathbf{y})$ be the indicator function of the complement of $A$. It follows that
\[
\mathcal{A}  = \int \int d\mathbf{x} d\mathbf{y} \left[\pi(\mathbf{x}, \mathbf{y}) I_A(\mathbf{x}, \mathbf{y}) + \pi(\mathbf{y}, \mathbf{x})I_{A^c}(\mathbf{x}, \mathbf{y})\right].
\]
Let $B$ be the set of points such that $\pi(\mathbf{x}, \mathbf{y}) = \pi(\mathbf{y}, \mathbf{x})$. Use of symmetry in Eq.~(\ref{eq:II4}) produces  $I_{A^c}(\mathbf{x}, \mathbf{y}) = I_A(\mathbf{y}, \mathbf{x}) + I_{B}(\mathbf{x}, \mathbf{y})$. As such, 
\begin{eqnarray}
\label{eq:II4a}
\nonumber
\mathcal{A}  = \int \int d\mathbf{x} d\mathbf{y} [\pi(\mathbf{x}, \mathbf{y}) I_A(\mathbf{x}, \mathbf{y}) + \pi(\mathbf{y}, \mathbf{x})I_{A}(\mathbf{y}, \mathbf{x}) \\
+ \pi(\mathbf{x}, \mathbf{y})  I_{B}(\mathbf{x}, \mathbf{y})]  = 2 \int \int d\mathbf{x} d\mathbf{y} \pi(\mathbf{x}, \mathbf{y}) I_A(\mathbf{x}, \mathbf{y}) \\ + \int \int d\mathbf{x} d\mathbf{y} \pi(\mathbf{x}, \mathbf{y}) I_{B}(\mathbf{x}, \mathbf{y}).
\nonumber
\end{eqnarray}
 Notice that $\pi(\mathbf{x},\mathbf{y})$ is symmetric if and only if $X(\mathbf{x}, \mathbf{y}) = 0$ and that $\pi(\mathbf{x}, \mathbf{y}) <  \pi(\mathbf{y}, \mathbf{x})$ if and only if $X(\mathbf{x}, \mathbf{y}) > 0$.  If follows that the acceptance probability $\mathcal{A}$ is the probability that $X(\mathbf{x}, \mathbf{y}) = 0$ plus twice the probability that $X(\mathbf{x}, \mathbf{y}) > 0$. That is, 
\begin{equation}
\label{eq:II2}
\mathcal{A} = P(X = 0) + 2P(X > 0). 
\end{equation}

For a system made up of a large number $n$ of independent and identically distributed subsystems sampled together,  Eq.~(\ref{eq:II2}) becomes
\begin{equation}
\label{eq:II9}
\mathcal{A}_n = P(S_n/n = 0) + 2P(S_n/n > 0). 
\end{equation}
Recall the definition of $S_n$ given by Eq.~(\ref{eq:II8}). The division by $n$ does not change the equalities or inequalities in Eq.~(\ref{eq:II2}) and is a formality that arranges Eq.~(\ref{eq:II9}) in the form typical of large deviation problems.

 By the strong law of large numbers, $S_n/n$ converges to minus the Kullback--Leibler divergence $D_1$ given by Eq.~(\ref{eq:I8}). If the detailed balance is not satisfied almost everywhere, then $-D_1 < 0$. Naturally, this implies $P(S_n/n = 0) \to 0$ and $P(S_n/n > 0) \to 0$, as $n \to \infty$. For large deviation problems, the decay of the last probabilities is exponentially fast. For the simple case discussed here, where $S_n$ is a sum of identical and statistically independent random variables, the law of the exponential decay is exactly known and is given by Cram\'er's theorem, the application of which leads again to
\begin{equation}
\label{eq:II10}
-\lim_{n \to \infty} n^{-1}\log(\mathcal{A}_n) = \frac{1}{2} D_{1/2}, 
\end{equation}
with $D_{1/2}$ defined by Eq.~(\ref{eq:I5}). The details of the proof are given in Appendix~II.

Eq.~(\ref{eq:II4a}) provides two different estimators for the acceptance probability, the fifty-fifty average of which is that implied by Eq.~(\ref{eq:I1}).  The first one is given by Eq.~(\ref{eq:II2}) and is only adequate for large values of the acceptance probability (larger than $1/2$). The second can be deduced by a similar reasoning, save for changing the sense of the inequality in the definition of the set $A$. It is given by
\begin{equation}
\label{eq:II4b}
\mathcal{A}  = P(X = 0) + 2\int \int_{X < 0} d\mathbf{x} d\mathbf{y} \pi(\mathbf{x}, \mathbf{y}) \exp[X(\mathbf{x}, \mathbf{y})].
\end{equation}
In words, we first test if the move is likely to be rejected. If the answer is positive, then we accumulate twice the ratio $\rho(\mathbf{y}) T(\mathbf{x}|\mathbf{y}) / \rho(\mathbf{x}) T(\mathbf{y}|\mathbf{x})$. We accumulate $1$ if detailed balance is already satisfied and $0$ otherwise. The second estimator is much better behaved than the first if the acceptance probability is small, which happens when many moves are likely to be rejected. In this case, the probability for the event $X < 0$ is large and we get adequate statistics. In fact, Eq.~(\ref{eq:II2}) implies $P(X > 0) \leq 1/2$, so that $P(X \leq 0) \geq 1/2$, meaning that the number of points necessary for the implementation of Eq.~(\ref{eq:II4b}) is always more than half the total number. In addition, the variance of the estimator is always more favorable, since  $\exp[X(\mathbf{x}, \mathbf{y})]$ is smaller than 1 whenever $X < 0$. In many cases, smaller means significantly smaller and the standard deviation of the estimator turns out to be roughly proportional to the value of the acceptance probability itself. 

The strategy presented in the preceding paragraph has been utilized before by the present author and collaborators in the context of replica exchange methods for the design of partition function estimators that alleviate the overlap problem \cite{Che06}. In fact, the source of the overlap problem is a poor statistics related to the fact that $P(X > 0)$ can be exponentially small. Motivated by these examples, we present and justify the general form of such estimators in Appendix~I.  

The practical relevance of Eq.~(\ref{eq:II4b}) is for tuning the various parameters controlling a simulation. There, good accuracy over few Monte Carlo cycles is required in conditions in which most attempted moves are rejected. On a computer, the equality $X = 0$ cannot be tested exactly in floating-point arithmetic, owing to inherent numerical errors. In addition, the estimating function implied by Eq.~(\ref{eq:II4b}) is not necessarily smaller than $1$ everywhere. As described in Appendix~I, an estimator that also addresses these issues is provided by
\begin{equation}
\label{eq:II4c}
e^{\min\{0, X(\mathbf{x}, \mathbf{y})\}} \times 
\left\{
\begin{array}{ll}
0, & \text{if} \; X(\mathbf{x}, \mathbf{y}) > \log(2), \\
1, & \text{if} \; |X(\mathbf{x}, \mathbf{y})| \leq \log(2), \\
2, & \text{if} \; X(\mathbf{x}, \mathbf{y}) < -\log(2).
\end{array}
\right.
\end{equation}

\section{Scaling of tuning parameters with the dimensionality: examples}

Eq.~(\ref{eq:I3}) produces the necessary and sufficient condition for the acceptance probabilities to remain finite upon large changes in various parameters, such as the dimensionality of the system. Let $D_{1/2}(d, \alpha)$ denote the R\'enyi divergence of order $1/2$ for a $d$-dimensional sampler utilizing a trial distribution additionally characterized by the parameter or the family of parameters $\alpha$. The necessary and sufficient condition for nonvanishing asymptotic acceptance probabilities is the existence of values $\alpha_d$ such that
\begin{equation}
\label{eq:III3}
D_{1/2}(d, \alpha_d) \leq M < \infty, \quad \forall \; d \geq 1.
\end{equation} 
In words, the sequence $D_{1/2}{(d, \alpha_d)}$ must be bounded. If $D_{1/2}{(d, \alpha_d)}$ has a subsequence increasing to infinity, then the corresponding acceptance probabilities will converge to zero, by the upper-bound inequality in Eq.~(\ref{eq:I3}). This establishes the necessity. Conversely, if $D_{1/2}{(d, \alpha_d)}$ is bounded from above by $M$, then the acceptance probabilities cannot fall below $e^{-M}/2$, by the lower-bound inequality. 

Albeit the index $d$ does not have to be the dimension, the dependence with the dimensionality is the natural problem to study, owing to the additivity of the entropy divergence for independent systems and trial distributions. The necessity for tuning the parameters $\alpha$ comes from an expected unlimited increase in the entropic divergence if these parameters are kept constant. For definiteness, assume that we are given a chemically homogeneous system. For proposals $T(\mathbf{y}, \mathbf{x})$ that are products of dimension-independent distributions, we expect the entropic divergence to increase linearly with the number of dimensions. Indeed, the proposal is already conditionally independent, whereas the particles making up a physical system must decorrelate in the thermodynamic limit. A better insight is provided by the Kullback-Leibler divergence, which, according to Eq.~(\ref{eq:II13a}), bounds the R\'enyi divergence of order $1/2$ from above. The behavior of $D_1$ closely mimics the behavior of the Boltzmann-Gibbs entropy except for the case where the proposal distribution introduces additional correlation. (Recall that the Boltzmann-Gibbs entropy and the Shannon entropy are given by the same formula, except for the multiplicative Boltzmann constant). Moreover, $D_1$ is easier to evaluate numerically if verification of linearity is desired in order to test for finite-size effects.

In the reminder of the section, we apply the condition to the random-walk Metropolis algorithm and the parallel tempering method in order to demonstrate the versatility of the entropic inequalities in predicting the scaling behavior in various situations. The random-walk Metropolis algorithm is constructed from a proposal distribution of the form 
\begin{equation}
\label{eq:III4}
T(\mathbf{x} + \mathbf{z}|\mathbf{x}) = {\alpha^{-1}} f(\alpha^{-1}\mathbf{z}).
\end{equation}
Here, $f(\mathbf{z})$ is a normalized distribution invariant under the transformation $\mathbf{z} \mapsto -\mathbf{z}$. For simplicity, we take $f(\mathbf{z})$ to be symmetric in each of the variables $z_i$. In addition, we require that $f(\mathbf{z})$ be a product of low-dimensional distributions, so that the entropic divergence does not increase faster than linearly. Typically, the scaling factor $\alpha > 0$ is tuned such that the acceptance probability lies somewhere between $20\%$ and $50\%$. The entropic divergence $D_{1/2}(d, \alpha)$ is
\begin{eqnarray*}
 - 2\log\left(\alpha^{-1}\int \! d\mathbf{x} \! \int \! d\mathbf{z} f(\alpha^{-1}\mathbf{z}) \left[\rho(\mathbf{x})\rho(\mathbf{x}+\mathbf{z})\right]^{1/2}  \right) \\
=  - 2\log\left(\int \! d\mathbf{x} \! \int \! d\mathbf{z} \left[\rho(\mathbf{x})\rho(\mathbf{x}+\alpha \mathbf{z})\right]^{1/2}  f(\mathbf{z}) \right).
\end{eqnarray*}
For given $d$, the entropic divergence converges to zero as $\alpha \to 0$. In the same limit, the trial distribution shrinks to a delta function. 

For small $\alpha$, the leading term in the entropic divergence, as follows from a Taylor expansion around $\alpha = 0$, is  the Fisher entropy. More precisely, we have
\begin{equation}
\label{eq:III5}
D_{1/2}(d, \alpha) = \frac{\alpha^2}{4}  \sum_{i = 1}^d \sigma_i^2 \int \frac{1}{\rho(\mathbf{x})}\left[\frac{\partial \rho(\mathbf{x})}{\partial x_i}\right]^2 d\mathbf{x}  + O(\alpha^4d), 
\end{equation}
where 
\[
\sigma_i^2 = \int  f(\mathbf{z})z_i^2 d\mathbf{z}, \quad 1 \leq i \leq d
\]
are the second-order moments of the kernel $f(\mathbf{z})$. Under the assumptions we made about $f(\mathbf{z})$, the first-order moments and the cross second-order moments are zero. The error has an implied linear-only dependence with the dimension owing to the assumed decorrelation of the particles in the thermodynamic limit. Eq.~(\ref{eq:III3}) implies that $\alpha_d$ must decrease as fast as $d^{-1/2}$ (more generally, as the inverse square root of the Fisher entropy). Since $\alpha_d \sigma_i$ is the standard deviation in the direction $i$ of the trial distribution, we see that the average displacements are proportional to the inverse square root of the number of dimensions  sampled together. That the scaling is correct has been  verified numerically for Lennard-Jones clusters \cite{Pre05} and found to hold true. In addition, the mathematical analysis performed by Roberts \emph{et al} \cite{Rob97} for independent systems leads to the same conclusion. 

The finitude requirement for the second-order moments of the function $f(\mathbf{z})$ sets some constraints on the length of the tails for proposal rules that are products of low-dimensional distributions. Normally, we would like to utilize such products in order to avoid introducing artificial correlation between distant particles. Also, functions $f(\mathbf{z})$ with long tails seem more adequate for potentials with rough topologies. A good compromise is provided by products of coordinate-scaled versions of the one-dimensional function
\begin{equation}
\label{eq:III5a}
f(z) = (1+z^2)^{-3/2} / 2.  
\end{equation}
Albeit $f(z)$ does not have a finite variance, its  second moment is only mildly divergent and the distribution seems appropriate up to a few tens of variables updated simultaneously. The advantage over a finite-variance function of the form $(1+z^2)^{-3/2-\epsilon}$ is that random numbers distributed according to $f(z)$ can be cheaply generated  by means of the identity $\zeta = (\xi - 1/2)/(\xi (1-\xi)^{1/2})$, with $\xi$ uniformly distributed on $(0, 1)$. The trial distribution just described appeared to be superior to the standard flat distribution in the Monte Carlo sampling of a path-integral implementation of the Onsager-Machlup formula \cite{Mil07}. Nevertheless, any clear advantage of a long-tail distribution is perhaps limited to rough potential surfaces. 

The second example we work out is for parallel tempering \cite{Liu01, Gey91}, where swaps are attempted between two large dimensional systems characterized by the inverse temperatures $\beta' \geq  \beta$ \cite{Pre04}.  The entropic divergence is given by 
\begin{equation}
\label{eq:III6}
D_{1/2}(d, R) = -2\log\left[\frac{Q(\bar{\beta})^2}{Q(\beta)Q(\beta')}\right],
\end{equation}
where $\bar{\beta} = (\beta + \beta')/2$ and $R = \beta'/\beta \geq 1$. Eqs.~10 and 20 of Ref.~\onlinecite{Pre04} lead to
\[
D_{1/2}(d, R) = 2\left(\frac{R-1}{R+1}\right)^2 C_V^{(d)}(\bar{\beta}) + O(d|R - 1|^3),
\]
where $C_V^{(d)}(\bar{\beta})$ is the potential contribution to the heat capacity of the physical system. 

Again, we can compensate the linear increase characteristic of the heat capacity by letting the ratio $R$ get close to 1 sufficiently fast. We must choose a dimension-dependent value $R_d$ such that
\begin{equation}
\label{eq:III7}
C_V^{(d)}(\bar{\beta})(R_d-1)^2 \leq M < \infty,  
\end{equation}
The last inequality predicts that the temperatures $\beta$ and $\beta'$ must get close one to another  as fast as $d^{-1/2}$, a condition that is also sufficient. That the heat capacity is the main cause for the decrease in the temperature ratios has been demonstrated before by Kofke \cite{Kof02}. The $d^{-1/2}$ scaling has been confirmed by numerous simulations \cite{Pre04, Fuk02} and also subjected to specific mathematical analysis \cite{Kof02, Pre04}.

\section{Analysis of the Smart Monte Carlo method}

 The force-biased Monte Carlo methods \cite{Pan78} attempt to improve their efficiency by sampling from a more suitable local distribution constructed with the help of the first-order derivatives of the potential. In one-dimensional notation, we have 
\begin{equation}
\label{eq:IV1}
e^{-\beta V(y)} = e^{-\beta V(x) - \beta V'(x) (y-x)} + O(|y-x|^2).
\end{equation}
The right-hand side must be multiplied by a windowing function that eventually determines the lengthscale on which the Taylor approximation is valid. The windowing function must also decay fast enough to infinity, so that to make the right-hand side integrable. It was originally taken to be constant on a symmetric interval centered about $x$. If, for ease of computation, it is taken to be a Gaussian, then
\[
e^{-\beta V(y)} = e^{-\beta V(x) - \beta V'(x) (y-x) - (y-x)^2/2\sigma^2} + O(|y-x|^2).
\]
Completing the square, we deduce that a suitable proposal distribution is 
\begin{equation}
\label{eq:IV2}
T(y|x) = \frac{1}{\sqrt{2\pi \sigma^2}} \exp\left\{- \frac{\left[y-x + \beta \sigma^2 V'(x)\right]^2}{2\sigma^2}\right\}.
\end{equation}
We notice that the trial move is the transition kernel for a Smoluchowski process (or generalized Brownian motion), and this is clearly true if we only care to set $\sigma^2 = 2Dt$, with $D$ being the diffusion constant and $t$ being the (short) transit time. Nevertheless, the equilibrium temperature is wrong, namely $\exp[-2\beta V(x)]$.  

The Smart Monte Carlo method was introduced by Rossky, Doll, and Friedman \cite{Ros78} not as a force-biased technique, but as a Metropolis-Hastings correction to a Smoluchowski process. The idea is that the latter process is already an ergodic sampler for the Boltzmann distribution. However, the short-time transition kernels that are numerically available are only asymptotically accurate. The transition kernel described by Eq.~(\ref{eq:IV2}), with the inverse temperature $\beta$  replaced by $\beta / 2$, is characteristic of the stochastic Euler's integrator. Rao and Berne \cite{Rao79} have subsequently modified the force-biased method and introduced a tuning parameter $\theta$ to control the force bias.   

We follow their approach and define a class of transition kernels depending on a parameter $\theta$ and having the form
\begin{equation}
\label{eq:IV3}
T_\theta(\mathbf{y}|\mathbf{x}) = \prod_{i = 1}^d \frac{1}{\sqrt{2\pi \sigma_i^2}} \exp\left\{- \frac{\left[y_i-x_i + \theta\beta \sigma_i^2 \nabla_iV(\mathbf{x})\right]^2}{2\sigma_i^2}\right\}.
\end{equation}
Notice that for multidimensional systems the transition kernel is a product of one-dimensional distributions. As discussed in the preceding section, we expect the entropic divergence to behave linearly in the thermodynamic limit, because the proposal does not introduce additional particle correlation over the Boltzmann distribution. Let us find the value of $\theta$ that maximizes the standard deviations $\sigma_i$ for large dimensional systems. In Appendix~III, we show that the relative entropy $D_{1/2}(\theta)$ is given by the formula
\begin{equation}
\label{eq:IV4}
D_{1/2}(\theta) = \beta^2\left(\theta-{1}/{2}\right)^2\left\langle\sum_{i=1}^d \left[\sigma_i \nabla_iV(\mathbf{x})\right]^2 \right\rangle + O(d\|\sigma\|^4).
\end{equation}
The brackets denote an average against the Boltzmann distribution and the quantity itself is again the Fisher entropy (of course, up to some scaling or multiplicative constants). 

Assume $\theta \neq 1/2$. From the asymptotic requirement
\[
 \beta^2\left(\theta-{1}/{2}\right)^2\left\langle\sum_{i=1}^d \left[\sigma_i \nabla_iV(\mathbf{x})\right]^2 \right\rangle \leq M < \infty, 
\]
we learn that the maximal standard deviations scale as
\begin{equation}
\label{eq:IV5}
\sigma_{i,d} \sim {\sigma_i^0}/{d^{1/2}}, 
\end{equation}
if the acceptance probability is to converge to a nonvanishing value in the limit $d \to \infty$.
Even more, the dominant term  in Eq.~(\ref{eq:IV4}) has the same value whether $\theta = 0$ or $\theta = 1$. It follows that, for sufficiently large systems, the original force-biased Monte Carlo method we considered in Eq.~(\ref{eq:IV2}) exhibits essentially the same average displacements as the unbiased method. 

The Smart Monte Carlo method corresponds to the choice of parameter $\theta = 1/2$, which is the only value that cancels the dominant Fisher entropy. As argued in Appendix~III, for $\theta = 1/2$, the decay of the entropic divergence is  as fast as $d\|\sigma\|^6$, but in general not faster. Again, the factor $d$ is always linear, albeit it needs not be so for systems that do not have a well-defined thermodynamic limit. The scaling of the standard deviations with the dimensionality improves to
\begin{equation}
\label{eq:IV6}
\sigma_{i,d} \sim {\sigma_i^0}/{d^{1/6}},
\end{equation}
where $\sigma_i^0$ are some asymptotic constants. This result has been obtained before by Roberts and Rosenthal \cite{Rob98} for independent systems and suggested by the work of Kennedy and Pendleton \cite{Ken91} on hybrid Monte Carlo.

 Incidently, the asymptotic expansion worked out in Appendix~III shows that the cancellation of the term $\|\sigma\|^4$ is due to the special values of the moments of the Gaussian window. The force-biased method can also achieve the $d^{-1/6}$ scaling provided that the original rectangular window is modified so that to match the first $5$ moments of a Gaussian. The simplest replacement is furnished by scaled products of 
\begin{equation}
(2\pi)^{-1/2}e^{-z^2/2} \approx \frac{2}{5}\delta(z) + \frac{3}{5} I_{[-\sqrt{5}, \sqrt{5}]}(z).
\end{equation}
The first term is Dirac's delta function, whereas the second is the indicator function for the interval $[-\sqrt{5}, \sqrt{5}]$. The utilization of the delta function does not pose programming problems. By its symmetry, the window is only part of the proposal step and conveniently cancels in the acceptance/rejection step, save for the normalization coefficient. A compactly supported window may alleviate some of the numerical issues associated with Euler's integrator, as the particles are prevented from moving arbitrarily far under harsh gradients. 

In the remainder of the section, as an illustration,  we determine numerically the scaling of the tuned standard deviations $\sigma_{i,d}$ for Lennard-Jones clusters. As well-known, the potential energy is given by
\begin{equation}
\label{eq:co12}
\mathrm{V_{tot}}(\mathbf{x}) = \sum_{i<j}^{N_p} \mathrm{V_{LJ}}(r_{ij})+\sum_{i=1}^{N_p}
\mathrm{V_c}(\mathbf{r}_i), 
\end{equation}
with $N_p$ being the number of particles. As usual, $\mathrm{V_{LJ}}(r_{ij})$ is the Lennard-Jones potential describing the interaction between the particles $i$ and $j$
\begin{equation}
\label{eq:co13}
\mathrm{V_{LJ}}(r_{ij}) = 4\epsilon_{LJ}\left
        [\left( \frac{\sigma_{LJ}}{r_{ij}}\right)^{12}
       -\left( \frac{\sigma_{LJ}}{r_{ij}}\right)^{6}\right].
\end{equation} 
$\mathrm{V_{c}}(\mathbf{r_i})$ is a confining potential of the form 
\begin{equation}
\label{eq:co14} 
\left\{
\begin{array}{l l}
0, & \text{if} \; \|\mathbf{r}_i-\mathbf{R_{cm}}\| < R_c, \\
5\cdot 10^3\epsilon_{LJ}\left({\|\mathbf{r}_i-\mathbf{R_{cm}}\|}/{R_c}\right)^{3}, & \text{otherwise}.
\end{array}
\right.
\end{equation}
Thus, the cluster is confined to its center of mass $\mathbf{R_{cm}}$. The values of the Lennard-Jones parameters $\sigma_{LJ}$ and $\epsilon_{LJ}$ used are 2.749 {\AA} and 35.6 K. They are characteristic of the Ne atoms. 

 \begin{figure}[!tbp] 
   \includegraphics[angle=270,width=8.0cm,clip=true]{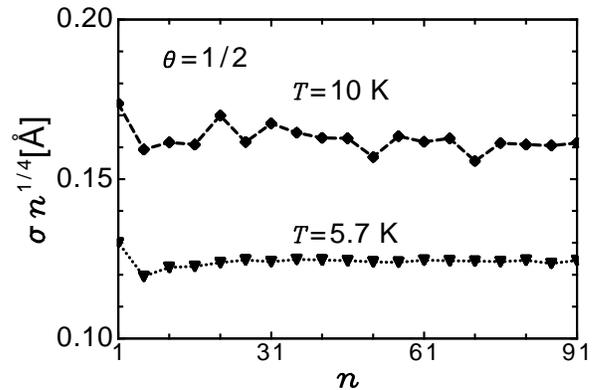} 
 \caption[sqr]
{\label{Fig:1}
Scaling of the standard deviations with the number of particles $n$ updated simultaneously for the $\mathrm{LJ}_{91}$ cluster at different low temperatures. A confining radius of $4\sigma_{LJ}$ is utilized. The updated particles are chosen randomly. The observed $n^{-1/4}$ behavior is caused by surface effects, which induce an artificial increase in the standard deviations for small $n$. The scaling is expected to improve to $n^{-1/6}$, in the thermodynamic limit.  
}
\end{figure}

The confining potential has been utilized before in Ref.~\cite{Mil07} for diffusion processes and found to be more adequate than a steeply-increasing high-order polynomial. The latter introduces numerical instabilities owing to the high gradients experienced by the atoms reaching the frontier. To overcome the tendency of SMC to stall if high gradients are met, we have coupled the sampler with a standard random-walk Metropolis algorithm. The latter has been randomly utilized $25\%$ of the time. The coupling with a random-walk Metropolis algorithm also serves a second purpose. Christensen et al \cite{Chr05} have pointed out that SMC is slow to attain stationarity if run by itself and that  an $n^{-1/4}$ scaling is more like to be observed in the transient regime. 

The typical strategy is to simulate a cluster having a large number of particles $N_p$, while utilizing updates of the whole system. In the second phase of the simulation, the underlying all-particle sampler is kept running. A smaller number of particles $n$ are attempted to be updated and the acceptance probability for such moves is accumulated by means of the estimator discussed in Section~III. Because the standard deviations for the proposal distribution keep changing, the attempted moves are not accepted, as this would alter the detailed balance. The computed acceptance probabilities are utilized to tune the standard deviations for the hypothetical $n$-particle sampler so that the final acceptance probability lies between $39\%$ and $41\%$. Since the moves are not accepted, we can choose those $n$ particles that are the closest to the center of mass. The coating provided by the exterior atoms minimizes the surface effects. Otherwise, if the $n$ particles are chosen randomly, then the standard deviations for small $n$ are artificially increased, owing to the larger mobility of the atoms at the surface. As shown in Fig.~\ref{Fig:1}, the surface effects create the appearance of an $n^{-1/4}$ scaling.

\begin{figure}[!tbp] 
   \includegraphics[angle=270,width=6.5cm,clip=true]{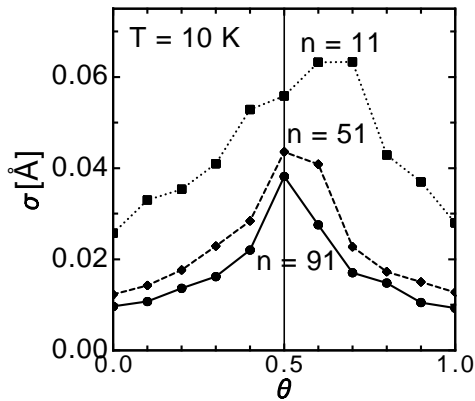} 
 \caption[sqr]
{\label{Fig:2}
Dependence of the standard deviations with the parameter $\theta$ for various numbers of particles updated simultaneously. The updated particles are those closest to the center of mass of a $500$-atom cluster. A confining radius of $5.5\sigma_{LJ}$ has been utilized. The $\theta$-dependence expressed by Eq.~(\ref{eq:IV4}) implies that the plotted curves become symmetric in the thermodynamic limit and feature a $|\theta - 1/2|^{-1}$ singularity.
}
\end{figure}

 Fig.~\ref{Fig:2} shows the common standard deviation of the trial distribution for different numbers $n$ of embedded particles that are updated simultaneously and for different values of the parameter $\theta$. As $n$ increases, the value $\theta = 1/2$ becomes the optimal one, in agreement with the theoretical predictions. For $n = 51$ and $n = 91$, it is also apparent that the standard deviations for the unbiased ($\theta = 0$) and fully biased ($\theta = 1$) Monte Carlo techniques are almost equal, again, in agreement with the theory.  For small $n$, the plots are biased toward larger $\theta$, which suggests that the guidance provided by the force bias is energetically favorable, yet eventually hindered by the entropic effects. 
 
\begin{figure}[!tbp] 
   \includegraphics[angle=270,width=8.5cm,clip=true]{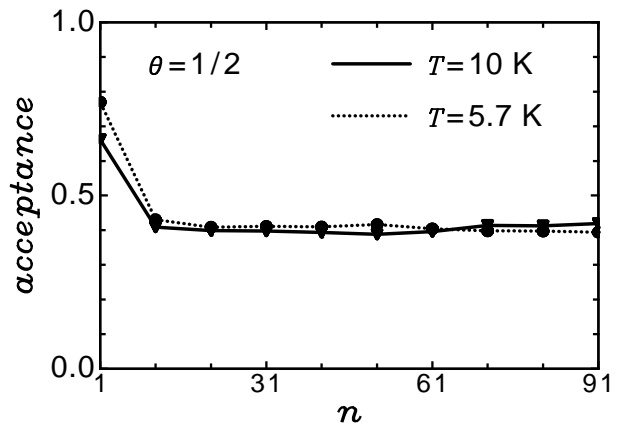} 
 \caption[sqr]
{\label{Fig:3}
Dependence of the observed acceptance probabilities with the number of particles $n$ updated simultaneously, for standard deviations scaled according to the $n^{-1/6}$ law. As for Fig.~\ref{Fig:2}, the updated particles are those closest to the center of mass of a $500$-atom cluster. In both cases, the Lennard-Jones potentials have been cut off beyond $3\sigma_{LJ}$. The error bars are given by the second significant digit. 
}
\end{figure}

Despite significant computational effort, we were not able to evaluate the standard deviations with sufficient accuracy for a proper demonstration of the $n^{-1/6}$ scaling. The reason is as follows. Even if we increase the number of particles updated simultaneously in an exponential fashion, say $n = 2^k$, the standard deviations for successive $n$ differ by a meager $12.2\%$. To correctly pinpoint the scaling, the relative error in the standard deviations needs to be about $2\%$. This is difficult to achieve by tuning. Many blocks are necessary, and the blocks must contain sufficiently many steps that the block averages for the acceptance probabilities have adequate statistical errors. However, a different approach is equally valid and computationally less expensive. We optimize the scaling parameters for the largest number of particles we plan to update (here, $n = 91$), and then decrease the number of atoms, say by $10$. For $n = 81$, the theory says that the acceptance probability remains constant if the standard deviations for $n = 91$ are increased by a factor of $[(n + 10) / n]^{1/6}$. The procedure is repeated down to $n=1$. To confirm the scaling, it suffices to check that the acceptance probabilities do remain constant. Since the standard deviations are kept constant for a given $n$, the many block estimates utilized for tuning can now be averaged to produce accurate estimates for the acceptance probabilities. The results presented in Fig.~\ref{Fig:3} adequately confirm the theoretical scaling.

\begin{figure}[!tbp] 
   \includegraphics[angle=270,width=8.5cm,clip=true]{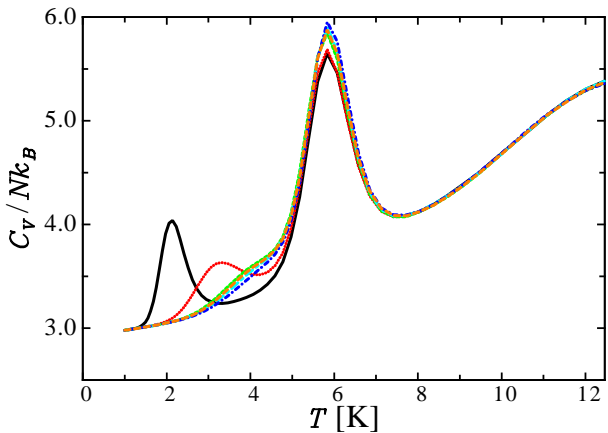}
 \caption[sqr] {\label{Fig:lj38pt} The evolution of the heat capacity
profiles is similar to that from Fig.~1 of Ref.~\onlinecite{Pre05b}. Each of the six curves has been
obtained by collecting  averages for groups of $175$
million steps. Early in the simulation, the heat capacity develops a
fake maximum, which is subsequently transformed in a ``shoulder.''  The last four curves fluctuate around this shoulder, with the fluctuations providing an estimate for the errors. }
\end{figure}

We have utilized the Smart Monte Carlo method to sample the LJ$_{38}$ cluster, which is a notoriously difficult problem, especially when a large confining radius $R_c = 3.612 \sigma_{LJ}$ is utilized. We have implemented SMC in the all-particle version and run it for a number of $1.25$ billion steps. The number of steps has been chosen so that the computational cost is the same as for the simulation performed in Ref.~\onlinecite{Pre05b}. There, a standard Metropolis sampler has been run for $360$ million sweeps, each sweep being composed of $38$ single-particle Metropolis moves. Even though parallel tempering is utilized, the LJ$_{38}$ cluster exhibits excessively long equilibration times. Heat capacities are overly sensitive to the lack of equilibration and serve as a good indicator for the progress of the simulation. The six curves shown in Fig.~\ref{Fig:lj38pt} have been evaluated by accumulating data over $175$ million SMC steps. They follow the same history as the curves presented in Fig.~1 of Ref.~\onlinecite{Pre05b}. Comparison shows that the Smart Monte Carlo method is marginally faster. Therefore, the simulation demonstrates the capability of the SMC method to sample difficult systems in the range of tens of atoms. For more complicated potential energies, the evaluation of the forces may be significantly faster than the evaluation of the potential differences for single particle moves, making the SMC method more advantageous (a non-trivial advantage is also the simplicity of the ensuing codes and the parallelization opportunity offered by the force-field). A more elaborate discussion is provided in the following section.

\section{Conclusions}

The present study allows us to make several general recommendations on the design of algorithms. We begin by considering the case of a single-particle sampling strategy. The approach can be regarded as a Monte Carlo implementation of a Gibbs sampler \cite{Liu01}. The latter is an idealized technique whereby the new coordinates for a particle are proposed directly from the conditional distribution of the particle. Obviously, a Gibbs sampler has a built-in correlation that sets a limit on the performance of Metropolis algorithms that update only a few variables at a time. As such, we cannot indefinitely improve the efficiency of the algorithms by  increasing the average displacements for the trial distribution. For example, in the case of a Lennard-Jones cluster, a particle essentially moves in a cage created by its environment. Even if we sample independently from the conditional distribution of the particle, we still propose moves mostly in this cage, which is the statistically important region.

Coping with the  Gibbs correlation requires updating more than one particle at a time. To set things in perspective, in the limit that all particles are updated and that the displacements are comparable with the size of the cluster, the sampling becomes almost independent. Unfortunately, by comparison with the Monte Carlo Gibbs-like sampler described in the preceding paragraph, the displacements are further decreased by the entropic effects. In the limit of small displacements, the Metropolis sampler is essentially diffusive. Consequently, if the decrease in the displacements is $n^{-\alpha}$, we will roughly need $n^{2\alpha}$ steps to cover distances comparable to those for a Gibbs-like sampler.

Despite its intrinsically local nature, an all-particle Metropolis sampler can be more efficient than a Gibbs sampler if the paths followed by the latter involve energetic barriers that are significantly larger than those characteristic of paths close to the minimum energy path. The latter paths are available to an all-atom sampler and more so to the Smart Monte Carlo method, whereas a Gibbs sampler may have little choice in effectuating a rare transition other than placing the particles successively on the edges of the conditional cages. In case of high cooperativeness for the minimum energy path \cite{Try04}, a large number of particles must be correctly positioned in order to cause a successful transition. Unfortunately, the transition probability is exponentially decreasing with the aforementioned number of particles, because probabilities for individual particles multiply. 

Which strategy must be utilized depends also on the computational cost to implement a given step. For general potentials, there might be the case that updating a single particle costs the same as updating all particles. If this is true, then it seems more advantageous to utilize the all-particle Metropolis algorithm. Indeed, we need $n$ calls to the potential function to defeat the entropic effects and span reasonable distances. The Gibbs-like sampler makes $n$ calls to the potential function to complete a sweep. However, the first technique is in principle not limited by the formation of cages. Nevertheless, in practice, the situation is more complicated because updating a single particle rarely requires a full evaluation of the potential. For example, a Lennard-Jones system can be updated one particle at a time in such a way that the cost for a complete sweep is twice the cost for an all-particle move. The gain in computational efficiency is large enough that the Monte Carlo implementation of the Gibbs sampler is the default sampler for Lennard-Jones clusters or liquids. Owing to its superiority in dealing with the entropic effects, the Smart Monte Carlo method is a good candidate for sampling subsystems containing tens or hundreds of atoms. 

To summarize, the ideal sampling strategy would divide the system in small subsystems (which may overlap) in a way that optimally compromises on the following requirements: i) the correlation between the subsystems be as small as possible and ii) the dimensionality of each subsystem be as small as possible. An efficient implementation may require the utilization of variables other than Cartesian ones, as the subsystems may become nearly uncorrelated in such coordinates. Each subsystem, which is described by strongly correlated degrees of freedom, is sampled utilizing an all-particle strategy or the Smart Monte Carlo method, if forces are easily computable. Finally, if techniques of reducing the dimensionality of the system are not available, then the designer should introduce tuning parameters that can arbitrarily increase the similarity between the direct and reverse sampling distributions. A case is constituted by the replica-exchange techniques. If tuning parameters are not provided, then one unavoidably ends up with an algorithm that  suddenly stops working for systems of sufficiently large dimensionality, owing to the severe exponential decay in the acceptance probability. 

The present study seems to favor coarse-graining and constrained-sampling as techniques capable of dealing with the negative impact of the dimensionality on the maximal displacements. Such approaches reduce the effective dimensionality  to that of the space of coarse variables, ideally described by their marginal distributions, or to that of the manifold defined by the set of constraints. Resolution exchange \cite{Lym06} can restore exactness in the first case, while at the same time improve the rare-event statistics, owing to the superior computational performance for the coarser levels of resolution. In this context, it appears that the estimators for the average acceptance probabilities derived here may prove useful in determining with reliability the degree of overlap between replicas of different resolutions.

Finally, on a separate topic, the influence of dimensionality on the global properties of molecular dynamics integrators is an issue that has to be more carefully investigated. As Kennedy and Pendleton have shown, Hybrid Monte Carlo, albeit guaranteeing exactness by means of detailed balance, requires time steps that decrease as a fractional power with the dimensionality of the system. The cause is precisely the entropic divergence documented here, this time between the phase-space Boltzmann distributions of the true and shadow Hamiltonians. Thinking that systems of $10^6$ degrees of freedom are typical nowadays, a decrease  in the time step by a factor of $10$ is not trivial. However, more worrisome seems to be the reverse scenario, where Hybrid Monte Carlo is not utilized. It is not at all clear whether or not the extensive entropic divergence may lead to differences between the exact and the integrator's statistics of such a nature that the validity of the latter as an approximation becomes questionable. Such a scenario could happen if the energy error is not equally divided between the degrees of freedom, and instead is transferred in an unfavorable way to crucial low-dimensional subsystems.   

\begin{acknowledgments} This work was supported in part by the National Science Foundation Grant No. CHE-0345280 and by the Director, Office of Science, Office of Basic Energy Sciences, Chemical Sciences, Geosciences, and Biosciences Division, U.S. Department of Energy under Contract No. DE-AC02-05CH11231. NERSC has provided the computing resources. 
\end{acknowledgments}

\appendix

\section{Low-variance estimators for strongly asymmetric distributions}

Sometimes, a sampling problem appears naturally in the form
\begin{equation}
\label{eq:AIII1}
\frac{\int\int d\mathbf{x}d\mathbf{y} f(\mathbf{x}, \mathbf{y})}{\int\int d\mathbf{x}d\mathbf{y} \pi(\mathbf{x}, \mathbf{y})},
\end{equation}
with $f(\mathbf{x}, \mathbf{y})$ symmetric and $\pi(\mathbf{x}, \mathbf{y})$ positive.  The distribution at the denominator is assumed unnormalized for the sake of generality. The default estimator $f(\mathbf{x}, \mathbf{y})/\pi(\mathbf{x}, \mathbf{y})$ might register large contributions to the variance arising from the pairs $(\mathbf{x}, \mathbf{y})$ where $\pi(\mathbf{x}, \mathbf{y})$ is small. This is apparent from the formula for the second moment, which is
\begin{equation}
\label{eq:AIII1a}
\frac{\int \int d\mathbf{x}d\mathbf{y} f(\mathbf{x},\mathbf{y})^2 \left/ \pi(\mathbf{x}, \mathbf{y}) \right.}{\int\int d\mathbf{x}d\mathbf{y}\pi(\mathbf{x}, \mathbf{y})}.
\end{equation}
 Nevertheless, for strongly asymmetric sampling distributions,  $\pi(\mathbf{y}, \mathbf{x})$ might be large for such pairs. It turns out that it is possible to construct estimators for which the variance behaves as if the sampling distribution were replaced by $\max\{\pi(\mathbf{x}, \mathbf{y}), \pi(\mathbf{y}, \mathbf{x})\}$  and which incur little penalty over the default estimator if the sampling distribution is not strongly asymmetric.  

Recall
\begin{equation}
\label{eq:AIII2}
X(\mathbf{x}, \mathbf{y}) = \log\left[{\pi(\mathbf{y}, \mathbf{x})}/{\pi(\mathbf{x}, \mathbf{y})}\right]
\end{equation}
and let $\epsilon \geq 0$. Define the sets 
\begin{equation}
\label{eq:AIII3}
A_\epsilon = \left\{(\mathbf{x}, \mathbf{y}) : X(\mathbf{x}, \mathbf{y}) < -\epsilon \right\}
\end{equation}
and
\begin{equation}
\label{eq:AIII4}
B_\epsilon = \left\{(\mathbf{x}, \mathbf{y}) : |X(\mathbf{x}, \mathbf{y})| \leq \epsilon \right\}.
\end{equation}
Notice that
\begin{equation}
\label{eq:AIII5}
I_{A_\epsilon^c}(\mathbf{x}, \mathbf{y}) = I_{A_\epsilon}(\mathbf{y}, \mathbf{x}) + I_{B_\epsilon}(\mathbf{x}, \mathbf{y}),
\end{equation}
a property that stems from the antisymmetry relation $X(\mathbf{x}, \mathbf{y}) = -X(\mathbf{y}, \mathbf{x})$.
Therefore, we can write
\begin{eqnarray}
\label{eq:AIII6}
\nonumber
\int\int d\mathbf{x}d\mathbf{y} f(\mathbf{x}, \mathbf{y}) = \int\int d\mathbf{x}d\mathbf{y} f(\mathbf{x}, \mathbf{y})I_{A_\epsilon}(\mathbf{x}, \mathbf{y}) \\ + \int\int d\mathbf{x}d\mathbf{y} f(\mathbf{x}, \mathbf{y}) I_{A_\epsilon}(\mathbf{y}, \mathbf{x}) \\ + \int\int d\mathbf{x}d\mathbf{y} f(\mathbf{x}, \mathbf{y})I_{B_\epsilon}(\mathbf{x}, \mathbf{y}).
\nonumber
\end{eqnarray}
The symmetry  of $f(\mathbf{x}, \mathbf{y})$ implies that the first two terms of the right-hand side are equal, so that
\begin{eqnarray}
\label{eq:AIII7}
\nonumber
\int\int d\mathbf{x}d\mathbf{y} f(\mathbf{x}, \mathbf{y}) = 2\int\int d\mathbf{x}d\mathbf{y} f(\mathbf{x}, \mathbf{y})I_{A_\epsilon}(\mathbf{x}, \mathbf{y}) \\  + \int\int d\mathbf{x}d\mathbf{y} f(\mathbf{x}, \mathbf{y})I_{B_\epsilon}(\mathbf{x}, \mathbf{y}).
\end{eqnarray}
This leads to the estimating function
\begin{equation}
\label{eq:AIII8}
\frac{f(\mathbf{x},\mathbf{y})}{\pi(\mathbf{x}, \mathbf{y})}
\times \left\{
\begin{array}{ll}
0, & \text{if} \; X(\mathbf{x}, \mathbf{y}) > \epsilon, \\
1, & \text{if} \; |X(\mathbf{x}, \mathbf{y})| \leq \epsilon, \\
2, & \text{if} \; X(\mathbf{x}, \mathbf{y}) < -\epsilon.
\end{array}
\right.
\end{equation}

The estimator given by Eq.~(\ref{eq:II4c}) is obtained by setting $\pi(\mathbf{x}, \mathbf{y})$ equal to $\rho(\mathbf{x})T(\mathbf{y}|\mathbf{x})$ and $f(\mathbf{x},\mathbf{y})$ equal to $\min\{\pi(\mathbf{x}, \mathbf{y}), \pi(\mathbf{y}, \mathbf{x})\}$. The partition function estimator given by Eq.~(12) of Ref.~\onlinecite{Che06} is obtained by setting $\pi(\mathbf{x}, \mathbf{y})$ equal to $\rho_i(\mathbf{x})\rho_j(\mathbf{y})$ and $f(\mathbf{x},\mathbf{y})$ equal to $\rho_j(\mathbf{x})\rho_j(\mathbf{y})$. 

In order to avoid a divergent variance caused by small values of $\pi(\mathbf{x}, \mathbf{y})$, it suffices to let $\epsilon$ be smaller or equal to  $\log(2)$. Indeed, on the set $|X(\mathbf{x}, \mathbf{y})| \leq \epsilon$, the inequality 
\[
\pi(\mathbf{y},\mathbf{x}) \leq e^{\epsilon} \pi(\mathbf{x},\mathbf{y}) \leq 2 \pi(\mathbf{x},\mathbf{y}) 
\]
implies 
\[
2\pi(\mathbf{x},\mathbf{y}) \geq \max\left\{\pi(\mathbf{x},\mathbf{y}), \pi(\mathbf{y},\mathbf{x})\right\}.
\]
On the other hand, on the set $X(\mathbf{x}, \mathbf{y}) < -\epsilon$, 
\[
\pi(\mathbf{x}, \mathbf{y}) = \max\left\{\pi(\mathbf{x},\mathbf{y}), \pi(\mathbf{y},\mathbf{x})\right\}.
\]
Therefore, the square of the estimating function is smaller than 
\begin{equation}
\label{eq:AIII9}
\frac{2f(\mathbf{x},\mathbf{y})^2}{\pi(\mathbf{x}, \mathbf{y})\max\{\pi(\mathbf{x}, \mathbf{y}), \pi(\mathbf{y}, \mathbf{x})\}}
\times \left\{
\begin{array}{ll}
0, & \text{if} \; X(\mathbf{x}, \mathbf{y}) > \epsilon, \\
1, & \text{if} \; |X(\mathbf{x}, \mathbf{y})| \leq \epsilon, \\
2, & \text{if} \; X(\mathbf{x}, \mathbf{y}) < -\epsilon,
\end{array}
\right.
\end{equation}
the expected value of which is
\begin{equation}
\label{eq:AIII10}
2\frac{\int \int d\mathbf{x}d\mathbf{y} f(\mathbf{x},\mathbf{y})^2 \left/ \max\{\pi(\mathbf{x}, \mathbf{y}), \pi(\mathbf{y}, \mathbf{x})\} \right.}{\int\int d\mathbf{x}d\mathbf{y}\pi(\mathbf{x}, \mathbf{y})},
\end{equation}
by the symmetry of the integrand in the numerator. The value of Eq.~(\ref{eq:AIII10}) can be at most twice larger than the second moment of the default estimator. However, it may actually be significantly smaller for strongly asymmetric distributions, for which $\max\{\pi(\mathbf{x}, \mathbf{y}), \pi(\mathbf{y}, \mathbf{x})\}$ may often be substantially larger than $\pi(\mathbf{x},\mathbf{y})$. Notice that Eq.~(\ref{eq:II2}) is still valid and implies $P(X \leq 0) \geq 1/2$. Thus, the penalty of at most $2$ has a statistical origin. For nearly symmetric distributions, we may end up utilizing only half (but not less) of the total number of points sampled.  

Among the possible values of $\epsilon \leq \log(2)$, the choice $\epsilon = \log(2)$ has one  additional desirable property. If the ratio $f(\mathbf{x}, \mathbf{y})/\pi(\mathbf{x}, \mathbf{y})$ is bounded everywhere by a constant $M > 0$, the modified estimator should ideally exhibit this property as well. An example is provided by the modified estimator for the acceptance probability, which, desirably, should be less or equal to $1$ everywhere. The choice $\epsilon = \log(2)$ guarantees the boundness property. The proof is as follows. The modified estimator can only be out of bounds on the set $X(\mathbf{x}, \mathbf{y}) < -\log(2)$, possibly because of the multiplicative factor of $2$. However, on this set, the inequality $\pi(\mathbf{y}, \mathbf{x}) < \pi(\mathbf{x}, \mathbf{y}) / 2$ holds. The boundness by $M$ then follows from the relations
\[
2\frac{\left|f(\mathbf{x}, \mathbf{y})\right|}{\pi(\mathbf{x}, \mathbf{y})} = 2\frac{\left|f(\mathbf{y}, \mathbf{x})\right|}{\pi(\mathbf{x}, \mathbf{y})} < \frac{\left|f(\mathbf{y}, \mathbf{x})\right|}{\pi(\mathbf{y}, \mathbf{x})} \leq M. 
\]

\section{Proof of Eq.~(\ref{eq:II10}) by Cram\'er's theorem}
As otherwise the statement of Eq.~(\ref{eq:II10}) is trivially true, we shall assume that the set of points for which the detailed balance condition is violated
\begin{equation}
\label{eq:AII1}
\rho(\mathbf{x})T(\mathbf{y}|\mathbf{x}) \neq  \rho(\mathbf{y})T(\mathbf{x} | \mathbf{y})
\end{equation}
has nonvanishing probability. The rejection mechanism in the Metropolis-Hastings method has the purpose of correcting the lack of detailed balance of the trial distribution.  
 Owing to Eq.~(\ref{eq:AII1}), we have $D_1 > 0$. Recall Eq.~(\ref{eq:II1}) and consider the moment generating function
\begin{equation}
\label{eq:AII2}
\phi(\theta) = \mathbb{E}\exp(\theta X) = \int \int \pi(\mathbf{x}, \mathbf{y}) \left[\frac{\pi(\mathbf{y}, \mathbf{x})}{\pi(\mathbf{x}, \mathbf{y})}\right]^\theta d\mathbf{x}d\mathbf{y}.
\end{equation} 
Observe that the first derivative
\[
\phi'(\theta) = \int \int \pi(\mathbf{x}, \mathbf{y}) \left[\frac{\pi(\mathbf{y}, \mathbf{x})}{\pi(\mathbf{x}, \mathbf{y})}\right]^\theta  \log\left[\frac{\pi(\mathbf{y}, \mathbf{x})}{\pi(\mathbf{x}, \mathbf{y})}\right] d\mathbf{x}d\mathbf{y}
\]
is defined on the interval $[0, 1]$. In fact $\phi'(0) = -D_1$, $\phi'(1) = D_1$, and $\phi'(1/2) = 0$. The last relation follows easily by symmetry. If $a$ is such that $-D_1 < a < D_1$, then the equation $a = \phi'(\theta_a)/\phi(\theta_a)$ has a unique solution in the interval $(0, 1)$, owing to the fact that $\phi'(\theta)/\phi(\theta)$ is increasing. The reader can look at the moment generating function and see that the derivative of $\phi'(\theta)/\phi(\theta)$ is a ``heat capacity.'' Moreover, Eq.~(\ref{eq:AII1}) demands that this heat capacity be continuous and non-zero on $(-D_1, D_1)$. From the implicit mapping theorem, it follows that the solution $\theta_a$ is continuous in $a$. Of note is that $\theta_{0} = 1/2$.

According to Cram\'er's theorem, which is stated at the end of the appendix, we have
\[
\lim_{n\to \infty}\frac{1}{n} \log P(S_n/n \geq a) \to -a\theta_a + \log \phi(\theta_a), 
\]
for all $a \in (-D_1, D_1)$.
If $a > 0$, then, by Eq.~(\ref{eq:II9}), 
\[
2P(S_n / n \geq 0) \geq \mathcal{A}_n \geq 2P(S_n/n \geq a), 
\]
from where it follows that
\begin{eqnarray*}&&
\log\phi(1/2) \geq \limsup_{n\to \infty}\frac{1}{n} \log \mathcal{A}_n \\ && \geq \liminf_{n\to \infty} \frac{1}{n} \log \mathcal{A}_n \geq -a\theta_a + \log \phi(\theta_a).
\end{eqnarray*}
Letting $a \to 0$, by means of the aforementioned continuity, we obtain the desired result, namely
\begin{equation}
\label{eq:AII3}
\lim_{n\to \infty}\frac{1}{n}\log \mathcal{A}_n = \log\phi(1/2) = -\frac{1}{2}D_{1/2} < 0.
\end{equation}
To see that $D_{1/2} > 0$, observe that the function $f(x) = x^{1/2}$ is concave. Jensen's inequality as applied to Eq.~(\ref{eq:AII2}) produces $\phi(1/2) \leq 1$, with equality if and only if $\pi(\mathbf{x}, \mathbf{y}) = \pi(\mathbf{y}, \mathbf{x})$ holds almost surely, which is contrary to our assumption. Nevertheless, if the last equality holds almost surely, then Eq.~(\ref{eq:II10}) is trivially true, since $\mathcal{A}_n = 1$ and $D_{1/2} = 0$.

To help the reader understand the proof, we give the statement of Cram\'er's large deviation theorem \cite{Dur96}.

\begin{1} 
Let $\{X_n; n \geq 1\}$ be a sequence of independent and identically distributed random variables and let $S_n = \sum_{i=1}^n X_i$ be the partial sums. Let $\mu = \mathbb{E}X_i$ and pick $a > \mu$. Suppose $\phi(\theta)=\mathbb{E}\exp(\theta X_i)$ is defined on the interval $[0, \theta_+)$ and that the distribution of $X_i$ is not a point mass. Assume also that there is $\theta_a \in (0, \theta_+)$ so that $a = \phi'(\theta_a)/\phi(\theta_a)$. Then as $n \to \infty$, 
\[
n^{-1}\log P(S_n \geq na) \to - a \theta_a + \log\phi(\theta_a). 
\]
\end{1}

\section{Proof of Eq.~(\ref{eq:IV4})}

 We first consider the case $d = 1$. Write $D_{1/2}(\theta)$ in the form $D_{1/2}(\theta) = -2\log S(\theta)$. Following the substitution $y = x + \sigma z$, $S(\theta)$ is given by the expression 
\begin{eqnarray}
\label{eq:AI1}
\nonumber
S(\theta) = \frac{1}{Q(\beta)\sqrt{2\pi}}\int_\mathbb{R}dx \int_\mathbb{R}dz e^{-\beta [V(x)+V(x+z\sigma)]/2}  \\ \times e^{-\left[z+\theta\beta \sigma V'(x)\right]^2/4} e^{-\left[z-\theta\beta \sigma V'(x+z\sigma)\right]^2/4}. 
\end{eqnarray}
By interchanging the order of integration and performing the substitution $x' = x + z\sigma/2$, $S(\theta)$ is seen to equal
\begin{eqnarray}
\label{eq:AI2}
\nonumber
\frac{1}{Q(\beta)\sqrt{2\pi}}\int_\mathbb{R}dx \int_\mathbb{R}dz e^{-\frac{\beta}{2} [V(x-z\sigma/2)+V(x+z\sigma/2)]}  \\ \times e^{-\left[z+\theta\beta \sigma V'(x-z\sigma/2)\right]^2/4} e^{-\left[z-\theta\beta \sigma V'(x+z\sigma/2)\right]^2/4}. 
\end{eqnarray}
The integrand in this equation is symmetric with respect to the variable $z$. Rearrangement of the exponent produces
\begin{eqnarray}
\label{eq:AI3}
\nonumber
S(\theta) &=& \frac{1}{Q(\beta)}\int_\mathbb{R}dx e^{-\beta V(x)} \frac{1}{\sqrt{2\pi}} \int_\mathbb{R}dz e^{-z^2/2} \\
\nonumber && \times e^{-\beta [V(x-z\sigma/2)+V(x+z\sigma/2)-2V(x)]/2}   \\
&& \times e^{\theta\beta z\sigma [V'(x+z\sigma/2)-V'(x-z\sigma/2)]/2} \\ 
 \nonumber && \times e^{-\theta^2\beta^2 \sigma^2 [V'(x-z\sigma/2)^2 + V'(x+z\sigma/2)^2]/4}. 
\end{eqnarray}

The reason we write $S(\theta)$ in this form is that $\sigma$ appears either in the form $\sigma^2$ or as a product $z\sigma$. Since the odd moments of the Gaussian measure vanish, it follows that the Taylor expansion at $\sigma = 0$ contains only terms of the form $\sigma^{2k}$. Evaluating the first two nonvanishing terms explicitly, we obtain as an intermediate result
\begin{eqnarray}
\label{eq:AI4}
\nonumber
S(\theta) = 1 - \frac{1}{Q(\beta)}\int_\mathbb{R}dx e^{-\beta V(x)} \left[\frac{\theta^2\beta^2\sigma^2}{2}V'(x)^2 \right. \\ \left. + \frac{\beta \sigma^2}{2}\left(\frac{1}{4}-\theta \right)V''(x) \right] + O(\sigma^4). 
\end{eqnarray}
Integration by parts produces
\[
\int_\mathbb{R}dxe^{-\beta V(x)} V''(x) = \int_\mathbb{R}dxe^{-\beta V(x)} \beta V'(x)^2.  
\]
Eq.~(\ref{eq:AI4}) reduces to 
\begin{equation}
\label{eq:AI5}
S(\theta) = 1 - \frac{\beta^2\sigma^2(\theta-1/2)^2 }{2Q(\beta)}\int_\mathbb{R}dx e^{-\beta V(x)}V'(x)^2 + O(\sigma^4). 
\end{equation}
Since  $D_{1/2}(\theta) = -2\log S(\theta)$, we obtain
\begin{equation}
\label{eq:AI6}
D_{1/2}(\theta) = \beta^2\sigma^2 (\theta-1/2)^2 \left\langle V'(x)^2\right\rangle_\beta + O(\sigma^4), 
\end{equation}
where we have utilized the bra-ket notation to denote thermodynamic averages for the Fisher entropy.  For the $d$-dimensional case, the reader can reduce the problem to the form specified by Eq.~(\ref{eq:AI3}), perform the expansion, and then notice that all averages involving cross terms $z_i z_j$ with $i \neq j$ vanish. One eventually obtains the form specified by Eq.~(\ref{eq:IV4}). 

For $\theta = 1/2$, we can actually show that the decay of the entropy divergence is faster than the quartic rate implied by Eq.~(\ref{eq:AI6}). The decay is as fast as $O(\|\sigma\|^6)$, but in general not faster. Expensive yet straightforward calculations show that the quartic term is given by the thermodynamic average of the sum over $1 \leq i,j \leq d$ of the terms
\begin{eqnarray}
\label{eq:AI7}
\nonumber
-\frac{\sigma_i^2\sigma_j^2}{64}\left\{ 
{3\beta} \partial_{iijj}V(\mathbf{x}) + \beta^2 \left[-2\partial_{ij}V(\mathbf{x})\partial_{ij}V(\mathbf{x}) \right. \right. \quad \\
\left. + \partial_{ii}V(\mathbf{x})\partial_{jj}V(\mathbf{x}) -4 \partial_{i}V(\mathbf{x})\partial_{ijj}V(\mathbf{x}) \right]
\qquad \\ \left.
- 2\beta^3 [\partial_{i}V(\mathbf{x})]^2\partial_{jj}V(\mathbf{x}) + \beta^4[\partial_{i}V(\mathbf{x})]^2[\partial_{j}V(\mathbf{x})]^2 \right\}.
\nonumber
\end{eqnarray}
Again, we utilize integration by parts to replace the integrands with equivalent ones. The goal is to remove the integrands containing high order derivatives. By means of the identities
\begin{eqnarray*}
&& 
\left\langle \partial_{iijj}V(\mathbf{x}) \right\rangle_\beta = \beta \left\langle \partial_{i}V(\mathbf{x})\partial_{ijj}V(\mathbf{x}) \right\rangle_\beta \\
&& = \beta^2 \left\langle [\partial_{i}V(\mathbf{x})]^2\partial_{jj}V(\mathbf{x})\right\rangle_\beta - \beta \left\langle\partial_{ii}V(\mathbf{x})\partial_{jj}V(\mathbf{x}) \right\rangle_\beta, 
\end{eqnarray*}
we can replace Eq.~(\ref{eq:AI7}) with
\begin{eqnarray*} &&
\frac{\sigma_i^2\sigma_j^2}{64}\left\{ 
2\beta^2\partial_{ij}V(\mathbf{x})\partial_{ij}V(\mathbf{x})  - 2\beta^2 \partial_{ii}V(\mathbf{x})\partial_{jj}V(\mathbf{x}) \right. \\ && \left.
+ 3\beta^3 [\partial_{i}V(\mathbf{x})]^2\partial_{jj}V(\mathbf{x}) - \beta^4[\partial_{i}V(\mathbf{x})]^2[\partial_{j}V(\mathbf{x})]^2 \right\}.
\end{eqnarray*}
Furthermore, the equality
\begin{eqnarray*}
\left\langle [\partial_{i}V(\mathbf{x})]^2\partial_{jj}V(\mathbf{x})\right\rangle_\beta = \beta \left\langle [\partial_{i}V(\mathbf{x})]^2[\partial_{j}V(\mathbf{x})]^2\right\rangle_\beta \\ -2 \left\langle \partial_{i}V(\mathbf{x})\partial_{ij}V(\mathbf{x})\partial_{j}V(\mathbf{x})\right\rangle_\beta \quad
\end{eqnarray*}
leads to the equivalent integrand
\begin{eqnarray*} &&
\frac{\sigma_i^2\sigma_j^2}{32}\left\{ 
\beta^2\partial_{ij}V(\mathbf{x})\partial_{ij}V(\mathbf{x})  - \beta^2 \partial_{ii}V(\mathbf{x})\partial_{jj}V(\mathbf{x}) \right. \\ && \left.
+ \beta^3 [\partial_{i}V(\mathbf{x})]^2\partial_{jj}V(\mathbf{x}) - \beta^3\partial_{i}V(\mathbf{x})\partial_{ij}V(\mathbf{x})\partial_{j}V(\mathbf{x}) \right\}. 
\end{eqnarray*}

Letting $\phi(x) = \exp[-\beta V(x)/2]$, it is straightforward to show that the term of order $\|\sigma\|^4$ is the sum over all $1 \leq i, j \leq d$ of the terms
\begin{equation}
\label{eq:AI8}
\frac{\sigma_i^2\sigma_j^2}{8Q(\beta)} \int_{\mathbb{R}^d} \left\{\partial_{ii}\phi(\mathbf{x})\partial_{jj}\phi(\mathbf{x}) - [\partial_{ij}\phi(\mathbf{x})]^2\right\} d\mathbf{x}.
\end{equation}
However, the terms are equal to zero because of the identities
\begin{eqnarray*}
\int_{\mathbb{R}^d} \partial_{ii}\phi(\mathbf{x})\partial_{jj}\phi(\mathbf{x})d\mathbf{x} = \int_{\mathbb{R}^d} \phi(\mathbf{x})\partial_{iijj}\phi(\mathbf{x})d\mathbf{x}\\ = \int_{\mathbb{R}^d} \partial_{ij}\phi(\mathbf{x})\partial_{ij}\phi(\mathbf{x})d\mathbf{x},
\end{eqnarray*}
which follow by integration by parts. 

We conclude that, if $\theta = 1/2$, the dominant term in the expansion of the divergence $D_{1/2}(\theta)$ in powers of $\sigma$ is $O(\|\sigma\|^6)$. For the  quadratic potential $V(x) = 2x^2$ and for $\beta = 1$, numerical results show that $D_{1/2}(\theta)/\sigma^6 \to 1$, as $\sigma \to 0$. Thus, in general, we cannot hope for a better scaling.

\end{document}